\documentclass[11pt]{article}
\usepackage{amsmath}
\usepackage{amssymb,latexsym}
\usepackage[dvips]{graphicx}
\usepackage{rotating}
\def\eqref#1{(\ref{#1})}

\newtheorem{theorem}{Theorem}
\newtheorem{rema}{Remark}[theorem]

\def\rembox{\rule{1mm}{2mm}}

\newtheorem{ex}{Example}[theorem]
\renewcommand{\atop}[2]{\genfrac{}{}{0pt}{}{#1}{#2}}

\newcommand{\Real}{\mathbb R}
\newcommand{\Z}{\mathbb Z}
\newcommand{\fracs}[2]{ {#1} /{ #2} }
\newcommand{\p}{\hat p}
\newcommand{\q}{\hat q}
\newcommand{\X}{\hat X}
\newcommand{\osp}{\mathfrak{osp}}
\newcommand{\mexp}[1]{e^{#1}}
\newcommand{\Tr}{\mathrm{\mathop{Tr}}}
\renewcommand{\H}{\hat H}
\renewcommand{\b}{\hat b}

\newcommand{\PRL}{{\it Phys.\ Rev.\ Lett.\ }}
\newcommand{\JPA}{{\it J.\ Phys.\ A: Math.\ Gen.\ }}
\newcommand{\JMP}{{\it J.\ Math.\ Phys.\ }}

\begin{document}
\begin{center}
	\textbf{\Large The Wigner distribution function for the one-dimensional
	parabose oscillator}\\[3mm]
	\textsc{E.\ Jafarov$^{\dag, \ddag}$, S.\ Lievens$^{\dag, \S}$ 
	and J.\ Van der Jeugt$^\dag$}\\[2mm]
$\dag$ Department of Applied Mathematics and Computer Science,\\
Ghent University, Krijgslaan 281-S9, B-9000 Gent, Belgium \\
Elchin.Jafarov@ugent.be
Stijn.Lievens@ugent.be, 
Joris.VanderJeugt@ugent.be \\
$^\ddag$ Institute of Physics, Azerbaijan National Academy of Sciences, \\ 
Javid Avenue 33, AZ1143, Baku, Azerbaijan \\
 ejafarov@physics.ab.az  \\
$^\S$ Institute of Mathematics, Statistics and Actuarial Science \\ 
University of Kent,   Cornwallis Building,  Kent CT2 7NF,  United Kingdom \\
S.F.Lievens@kent.ac.uk
\end{center}


\begin{abstract}
In the beginning of the 1950's, Wigner introduced a fundamental deformation 
from the canonical quantum mechanical harmonic oscillator, which is nowadays 
sometimes called a Wigner quantum oscillator or a parabose oscillator.  Also, in quantum 
mechanics the so-called Wigner distribution is considered to be the closest
quantum analogue of the classical probability distribution over the phase space.
In this article, we consider which definition for such distribution function
could be used in the case of non-canonical quantum mechanics.  We then explicitly 
compute two different expressions for this distribution function for  the case of the 
parabose oscillator. Both expressions turn out to be multiple sums involving 
(generalized) Laguerre polynomials.  Plots then show that the Wigner distribution
function for the ground state of the parabose oscillator is similar in behaviour
to the Wigner distribution function of the first excited state 
of the canonical quantum oscillator.
\end{abstract}

\noindent {\em Running head:}  Wigner d.f.~for the parabose oscillator
\vskip 1cm

\section{Introduction}

The harmonic oscillator is one of the most appealing models due to its well
known solutions in classical mechanics and in quantum mechanics, and due to a
wide variety of applications in various branches of physics~\cite{moshinsky}.
In classical mechanics its solution is unique, and so it is in the canonical
non-relativistic treatment in quantum mechanics~\cite{landau}.  In this quantum
approach, the position and momentum operators ($\hat p$ and $\hat q$,
respectively) satisfy the canonical commutation relation $[ {\hat p,\hat q} ] =
- i\hbar $, and the model is described by its Hamiltonian $\hat H$ and the
Heisenberg equations.  This algebraic description can be translated to a
description by means of wave functions.  In the position representation, for
example, the operator $\hat q$ corresponds to multiplying functions by the
variable $q$ and the Heisenberg operator equations correspond to
Schr\"odinger's equation, the solutions of which are expressed through Hermite
polynomials. These solutions or wave functions lead to important position
probability functions.

To study the applicability of a system modelled by a harmonic oscillator, one
often turns to expressions for the joint quasi-probability function of momentum and
position, i.e.\ to a quantum analogue of phase space.  The Wigner distribution
function (d.f.)~\cite{wigner32} is the main theoretical tool here, as it is
considered as the closest quantum analogue of a classical distribution function
over the phase space.  The quantum harmonic oscillator is one of the first
systems for which the Wigner d.f.\ was computed analytically.  Nowadays, there
are also advanced experimental techniques that enable one to measure phase
space densities for certain quantum systems~\cite{ourjoumtsev, leibfried}, 
and thus compare theoretical
results with experimental results~\cite{smithey, lvovsky,mcmahon}.

It is in this context that certain deformations or generalizations of the
quantum harmonic oscillator are of relevance.  In particular, deformations that
are still exactly solvable and for which the Wigner d.f.\ can be computed
analytically are highly interesting.  One such model is the $q$-deformed
quantum harmonic oscillator, being exactly solvable~\cite{iwata, macfarlane, kagramanov}
and having found applications in various physical models~\cite{bonatsos}. 
Its Wigner d.f.\ has been computed in~\cite{jafarov}.

A deformation or generalization of the canonical quantum harmonic oscillator
that is more fundamental in nature was proposed by Wigner~\cite{wigner50}.  It
is known as the one-dimensional Wigner quantum oscillator or as the parabose
oscillator.  In Wigner's approach, the canonical commutation relations are not
assumed.  Instead of that, the compatibility of Hamilton's equations and the
Heisenberg equations is required.  This approach has been the fundamental
starting point of so-called Wigner Quantum Systems~\cite{Palev86}.  In such an
approach, one typically obtains several solutions (several algebraic
representations), of which only one corresponds to the canonical solution. It
is thus a natural and fundamental extension of canonical quantum mechanics.  It
is also of growing importance as deviations from the canonical commutation
relations (in particular of commuting position operators in the context of
non-commutative quantum mechanics)  are supported by investigations from
several field~\cite{Connes, Garay, Nair, Jackiw}.

The one-dimensional Wigner quantum oscillator itself was already solved by
Wigner~\cite{wigner50}, leading to an oscillator energy spectrum similar to the
canonical one but with shifted ground state energy equal to some arbitrary
positive number $a$.  In a sense, this number $a$ can be considered as the
deformation parameter, and the solutions of the Wigner quantum oscillator
reduce to the canonical one when $a=1/2$.  Wigner's formulation was also the
inspiration for the introduction of parabosons (and parafermions), as the
common boson commutator relation of the canonical case is, in Wigner's
approach, replaced by a triple relation that is also the defining relation of a
paraboson~\cite{green53}.  Later on, one realized that Lie superalgebras are the
natural framework for this algebraic description~\cite{ganchev}, and in particular
that the representations of the Wigner quantum oscillator are in one-to-one
correspondence with the unitary irreducible lowest weight representations of
the Lie superalgebra $\osp(1|2)$.

So the Wigner quantum oscillator, being the generalization of the canonical
one, could play an important role to explain various phenomena more precisely,
and its importance is undeniable.  However, so far its application has been
restricted in comparison with the canonical harmonic oscillator due to the fact
that its behaviour in phase space has not been studied.  In other words, there
is no expression of the joint quasi-probability distribution function (Wigner d.f.)
of position and momentum available for the Wigner quantum oscillator.  The main
problem is that the treatment a Wigner d.f.\ usually starts from the assumption
that the commutator of $\hat p$ and $\hat q$ is a constant, which is no longer
the case for the non-canonical Wigner quantum oscillator.  It is precisely this
problem which is being tackled and solved here.

In this paper, we first recall some basic facts of the one-dimensional Wigner
quantum oscillator or parabose oscillator. The emphasis is on the algebraic
description, in terms of the Lie superalgebra $\osp(1|2)$ and representations
characterized by a positive real number~$a$ (section~2).  In section~3 we
discuss the main definition of the Wigner d.f.\ in the current non-canonical
setting.  The main part of the paper is devoted to computing the Wigner d.f.\
$W_n(p,q)$ analytically, for the parabose oscillator in the $n$th excited pure
state $|n\rangle$.  For $n=0$ (and $n=1$), the calculation is somewhat simpler
than in the general case, and this is treated first in section~4. The
calculation involves various computational tools.  In some steps, series are
term wise integrated, so one should pay attention to convergence. Here, we
shall often assume that the parameter $a$ is of the form $a=1/2+m$, with $m$ a
nonnegative integer; in that case all series appearing are finite and there are
no convergence problems.  The final expression of $W_0(p,q)$ is essentially a
series in terms of Laguerre polynomials, depending only upon $p^2+q^2$, or
alternatively a confluent hypergeometric series in $p^2+q^2$, see~\eqref{W0-m}.
Section~5 then deals with the general case: the computation of $W_n(p,q)$.
Here, the analysis is quite involved, and some intermediate calculations are
performed in an Appendix B.  We obtain two alternative expressions for
$W_n(p,q)$, \eqref{A29} and~\eqref{A31}, both as series of Laguerre polynomials
in $p^2+q^2$.  Section~6 deals with some plots of the newly obtained Wigner
distribution functions, discusses some properties and summarizes some
conclusions.

To end the introduction, let us introduce some of the notation and 
classical results we are
going to use.  Many of the formulas encountered in this article will
involve binomial coefficients, factorials etc.  It is therefor convenient
to use the notation for (generalized) hypergeometric series~\cite{Slater}. 
Let $r$ and $s$
be nonnegative integers, a hypergeometric series with $r$ numerator
and $s$ denominator parameters is then defined as
\begin{equation}
{}_rF_s\left(\atop{a_1,\ldots,a_r}{b_1,\ldots, b_s} ;z\right) = 
\sum_{k\geq0} \frac{(a_1,\ldots,a_r)_k}{(b_1,\ldots,b_s)_k} \frac{z^k}{k!},
	\label{hyper}
\end{equation}
where
\begin{equation}
(a_1,\ldots,a_r)_k = (a_1)_k(a_2)_k\cdots (a_r)_k,\
\text{and}\ (a)_k = \frac{\Gamma(a+k)}{\Gamma(a)} = a(a+1)\cdots(a+k-1)
	\label{poch}
\end{equation}
is the rising factorial or Pochhammer symbol.  When one of the numerator
parameters $a_j$ is a negative integer, the series is terminating and in
fact a polynomial in $z$.  Of course, if one of the denominator parameters
is also a negative integer, it has to be \lq\lq more negative\rq\rq\ such
that the numerator vanishes before the denominator does.

All classical orthogonal polynomials can be written using this hypergeometric
notation.  In particular one has for the Laguerre polynomials~\cite{KoeSwart}:
\begin{equation}
	L_n^{(\alpha)}(z) = \frac{(\alpha+1)_n}{n!} \, 
	{}_1F_1\left( \atop{-n}{\alpha+1} ;z\right) = 
	\frac{(\alpha+1)_n}{n!}\, {}_1F_1(-n; \alpha+1;z).
	\label{Lag}
\end{equation}
Here, we also introduced an alternative notation for the confluent 
hypergeometric series ${}_1F_1$.  If the parameter $\alpha$ is $0$, we 
will simply write $L_n(z)$ instead of $L_n^{(0)}(z)$. Often, the $L_n(z)$ 
are called Laguerre polynomials, while the $L_n^{(\alpha)}$ are
called the generalized Laguerre polynomials.  For the classical 
orthogonality of the Laguerre polynomials to hold $\alpha$ needs to be
bigger than $-1$.  In particular, if $\alpha$ is a negative integer, the
definition~\eqref{Lag} is no longer valid since one has a negative integer
appearing as a denominator parameter.  The Laguerre polynomials can
be written in an alternative way such that they are defined for all values
of $\alpha$:
\begin{equation}
	L_n^{(\alpha)}(z) = \frac{1}{n!} \sum_{k=0}^n 
	\frac{(-n)_k(\alpha+k+1)_{n-k}}{k!}z^k.
	\label{Lag-gen}
\end{equation}

Many well known summation formulas can be very neatly expressed using the hypergeometric
notation.  Most notable are the binomial summation theorem:
\begin{equation}
	{}_1F_0({a};{\ };z) = (1-z)^{-a},
	\label{bin}
\end{equation}
with $|z| < 1$ if $a$ is not a negative integer,
and the Chu-Vandermonde summation theorem for a terminating Gauss hypergeometric
series:
\begin{equation}
	{}_2F_1\left( \atop{-n,a}{c};1 \right) = \frac{(c-a)_n}{(c)_n},
	\label{Vander}
\end{equation}
with $n$ a nonnegative integer.  There exists also a very extensive theory
of hypergeometric transformations, a very important transformation 
being Kummer's transformation for confluent hypergeometric series:
\begin{equation}
	{}_1F_1(b;c;z) = e^{z} \,  {}_1F_1(c-b;c;-z).
	\label{Kummer}
\end{equation}

\section{The one-dimensional parabose oscillator}

In this paragraph we briefly describe the treatment of a harmonic oscillator
as a Wigner quantum system.  Although first introduced by Wigner almost 60 
years ago, and described in some articles and book chapters, it may still 
be useful to recall the main results. 

Consider the Hamiltonian for a one-dimensional harmonic oscillator (in units with mass
and frequency both equal to 1):
\begin{equation}
\label{H}
\hat H = \frac{\p^2}{2} + \frac{\q^2}{2},
\end{equation}
where $\p$ and $\q$ denote respectively the momentum and position operator 
of the system.  Wigner \cite{wigner50} already noted that there are other 
solutions besides the canonical one if one only requires the compatibility
between the Hamilton and the Heisenberg equations. Thus, one drops the canonical
commutation relation $[\p,\q] = -i$  (we have taken $\hbar = 1$), and instead
one imposes the equivalence of Hamilton's equations
\begin{equation*}
\dot \p = -\frac{\partial H}{\partial q},\quad
\dot \q = \frac{\partial H}{\partial p}
\end{equation*}
and the Heisenberg equations
\begin{equation*}
\dot \p = i[\H, \p],\quad 
\dot \q = i[\H, \q].
\end{equation*}
Expressing this equivalence leads to the following compatibility conditions:
\begin{equation}
\label{CCs}
[\H, \p] = i\q,\quad
[\H, \q] = -i\p.
\end{equation}
So, one has to find operators $\p$ and $\q$, acting in some Hilbert space,
such that the compatibility conditions~\eqref{CCs} hold, with $\H$ given by~\eqref{H}.
Also, in this Hilbert space $\p$ and $\q$ have to be self-adjoint.

The solutions to~\eqref{CCs} can be found by introducing two new (yet unknown)
operators $\b^+$ and $\b^-$:
\begin{equation}\label{bpm}
\b^\pm = \frac{1}{\sqrt{2}}(\q \mp i\p), 
\end{equation}
or equivalently
\begin{equation*}
\q = \frac{1}{\sqrt{2}}(\b^+ + \b^-),\quad
\p = \frac{i}{\sqrt{2}}(\b^+ - \b^-).
\end{equation*}
It is then easily checked that
\begin{equation}\label{Hb}
\H = \frac12\{\b^-,\b^+\},
\end{equation}
and that the compatibility conditions~\eqref{CCs} are equivalent with
\begin{equation}
	[\{\b^-,\b^+\}, \b^\pm] = \pm 2 \b^\pm.
	\label{CCs2}
\end{equation}

The relations~\eqref{CCs2} are in fact the defining relations of 
one pair of parabose operators $\b^\pm$~\cite{green53}.
Moreover, from the self-adjointness of the position and momentum operators it 
follows that
\begin{equation}
(\b^\pm)^\dagger = \b^\mp.
	\label{bpm-adjoint}
\end{equation}
It is known that the Lie superalgebra generated by two odd elements 
$\b^\pm$ subject to the restriction~\eqref{CCs2} is the Lie superalgebra
$\osp(1|2)$~\cite{ganchev}.  
So, the solutions to our problem are given by the star representations
of the Lie superalgebra $\osp(1|2)$.  These are known, and are characterized
by a positive real number $a$ and a vacuum vector $|0\rangle$, such that
\begin{equation*}
\b^-|0\rangle = 0,\quad
\{\b^-,\b^+\} |0\rangle = 2a |0\rangle.
\end{equation*}
The representation space can then be shown to be the Hilbert space $\ell^2(\Z_+)$
with orthonormal basis vectors $|n\rangle$ ($n\in \Z_+$) and with the following 
actions:
\begin{equation}
\begin{aligned}
	\b^+ |2n\rangle & = \sqrt{2(n+a)}\,|2n+1\rangle, & & \quad &
	\b^- |2n\rangle & = \sqrt{2n}\,|2n-1\rangle, \\ 
	\b^+ |2n+1\rangle & = \sqrt{2(n+1)}\,|2n+2\rangle, & &  \quad & 
	\b^- |2n+1\rangle & = \sqrt{2(n+a)}\,|2n\rangle,  
\end{aligned}
	\label{bpm-actions}
\end{equation}
from which it immediately follows that
\begin{equation}
\{ \b^-, \b^+ \}|n\rangle = 2(n+a)\,|n\rangle.
	\label{action-anticomm}
\end{equation}

Using the actions~\eqref{bpm-actions}, one can now compute the action of
the commutator $[\p,\q]$:
\begin{equation}
[\p,\q]|2n\rangle = -2ai|2n\rangle ,\quad
[\p,\q]|2n+1\rangle = -2(1-a)i|2n+1\rangle.
	\label{action-comm-pq}
\end{equation}
{}So the canonical commutation relations are only satisfied for $a=1/2$ and
any representation with $a\neq 1/2$ thus leads to a 
non-canonical solution.  It is also important to note that for such non-canonical
solutions, the commutator $[\p,\q]$ is not constant.  This last fact has a serious
impact on the computation of exponential operators involving linear combinations
of $\p$ and $\q$ since the Baker-Campbell-Hausdorff theorem does not give 
rise to a simple formula.

The energy spectrum 
is now very easy to determine as well using~\eqref{action-anticomm} and~\eqref{Hb}:
\begin{equation*}
\H |n\rangle = (n+a)\,|n\rangle,
\end{equation*}
so one has an equidistant energy spectrum with ground level given by $a$. 
This once again confirms that only $a=1/2$ yields the canonical solution.

Also the wave functions for the parabose oscillator in the position and 
momentum representation are known.   Working in the position representation,
the operator $\q$ is still represented by \lq\lq multiplication by $q$\rq\rq,
the operator $\p$ however is no longer given by its canonical realization
$-i\frac{d}{dq}$, but instead has a different realization with an extra term 
(depending on $a$) coming in.  Using this realization the time-independent 
Schr\"odinger equation can then be solved yielding the following expressions
for the orthonormal wave functions in terms of 
Laguerre polynomials~\cite[Chapter 23]{ohnuki} and~\cite{Mukunda}:
\begin{equation}
\begin{split}
	\Psi _{2n}^{(a)}(q) & = (-1)^n  \sqrt {\frac{n!}{\Gamma( n + a) } }\, 
	|q|^{a-1/2}\,  e^{-q^2/2} L_n^{(a-1)}(q^2) \\
\Psi _{2n+1}^{(a)}(q) & = (-1)^n  \sqrt {\frac{n!}{\Gamma( n + a+1) } }\,  
|q|^{a-1/2}\, e^{-q^2/2} q L_n^{(a)}(q^2).
\end{split}
\label{wave}
\end{equation}
Again, for $a=1/2$, one recovers the very well known expression for the 
wave functions for a harmonic oscillator in terms of Hermite polynomials,
since there are formulas relating $L_n^{(\mp 1/2)}(x^2)$ to $H_{2n}(x)$ and
$H_{2n+1}(x)$ respectively.

\section{Distribution function for a non-canonical quantum system}

A major advantage of the phase-space formulation of quantum mechanics is 
the fact than one is no longer dealing with operators, but instead with
constant-number equations.  More in particular, the expectation value of
an arbitrary operator $\hat A(\p, \q)$ can be calculated using the distribution
function  $\Phi(p,q)$ as follows:
\begin{equation}\label{comp-av}
\Tr\{\hat \rho(\p,\q) \hat A(\p, \q) \} = \iint A(p,q) \Phi(q,p)\, dq\,dp,
\end{equation}
with $A(p,q)$ the scalar function obtained by replacing the operators $\p$
and $\q$ in the expression for $\hat A(\p, \q)$ by scalar variables $p$ and $q$.
Herein, $\hat \rho$ is the density operator, representing the state
of the system.  Also, both integrations run from $-\infty$ to $+\infty$; 
this will be the case for all integrations in this article.

It is also known that the distribution function $\Phi$ depends on the so-called 
correspondence rule used, this is, on the way one associates a function of 
non-commuting operators (thus an operator) to a given function of scalar variables.
A very popular association rule is the Weyl correspondence, yielding as
distribution function the so-called Wigner distribution function.  Very often, 
the definition for the Wigner d.f.~is written as follows:
\begin{equation}
W(p,q) = \frac{1}{2\pi} \int  
\langle q+\frac{\eta}{2} |\hat \rho |  q-\frac{\eta}{2} \rangle e^{-i\eta p} d\eta.
	\label{Wigner-df-class}
\end{equation}
This definition, however, is completely unsuitable for the case of non-canonical
quantum mechanics, since it relies heavily on the canonical commutation relations.
Indeed, it was shown by Tatarskii~\cite{tatarskii} (see also section~2.1 of Lee's review
paper~\cite{lee}) that the Wigner d.f.\ defined by first principles (see equation~\eqref{W_F} here)
gives rise to expression~\eqref{Wigner-df-class} only if $\p$ and $\q$ satisfy
$[\p,\q] = -i$. 
So if the canonical commutation relations are not satisfied, one should go back
to use this first principles approach, which is described in the rest of this section.

Another approach is the following~\cite{tatarskii}: is it clear that in order to define 
an association rule between functions of scalar variables $f(p,q)$ and
operators $\hat f(\p,\q)$ it suffices to define this correspondence 
for the function $F(\lambda, \mu) = \exp(i(\lambda p + \mu q))$, with
$\lambda$ and $\mu$ real variables.  Indeed, an arbitrary analytical function 
$f(p,q)$ can be obtained from $F(\lambda,\mu)$ in the following way:
\begin{equation*}
	f(p,q) = f(\frac{1}{i}\frac{\partial}{\partial \lambda},
	\frac{1}{i}\frac{\partial}{\partial \mu} ) 
	F(\lambda,\mu) \left|_{\lambda=\mu = 0}\right. . 
\end{equation*}
Note that this expression is well-defined since the partial derivate operators commute.
The operator corresponding to the function $f(p,q)$ under 
the chosen association rule, i.e.~for the particular choice of $\hat F$, is then 
given by
\begin{equation*}
	\hat f(\p,\q) = f(\frac{1}{i}\frac{\partial}{\partial \lambda},
	\frac{1}{i}\frac{\partial}{\partial \mu} ) 
	\hat F(\lambda,\mu)\left|_{\lambda=\mu = 0}\right..
\end{equation*}

Once a choice for $\hat F$ is made, the corresponding distribution function $W_F$
may be calculated as follows:
\begin{equation}
W_F(p,q) = \frac{1}{4\pi^2} \iint \langle \psi | \hat F(\lambda, \mu) | \psi \rangle
\mexp{-i(\lambda p + \mu q)} d\lambda d\mu,
	\label{W_F}
\end{equation}
where $\psi$ denotes the state of the system.  This definition is independent of 
the canonical commutation relations and still allows the computation of averages
by use of~\eqref{comp-av}.

We now make the following choice for $\hat F$~\cite{tatarskii, lee}:
\begin{equation}
	\hat F(\lambda,\mu) = \mexp{i(\lambda \p + \mu \q )} = \mexp{i\X},
	\label{choice-F}
\end{equation}
where we have introduced the operator $\X$:
\begin{equation}
	\X =  \lambda \p + \mu \q.
	\label{X}
\end{equation}

Since this is the only correspondence rule we are using in this article,
the subscript $F$ will be dropped from the notation.  On the other 
hand, as we will be concentrating on calculating explicit expressions
for the distribution function $W$ for the pure states $|n\rangle$, we will
introduce the subscript $n$ to indicate this:
\begin{equation}
	W_n(p,q) = \frac{1}{4\pi^2} \iint \langle n | \mexp{i\X} | n \rangle
	\mexp{-i(\lambda p + \mu q)} d\lambda d\mu.
	\label{W_n}
\end{equation}

Since in this article our goal will be to compute an explicit expression
for~\eqref{W_n}, it is convenient to recall the expression for the 
Wigner d.f.~$\tilde W_n$ in the case of canonical commutation relations:
\begin{equation}
	\tilde W_n(p,q) = \frac{(-1)^n}{\pi}\mexp{-p^2-q^2}\,L_n(2p^2+2q^2),
	\label{tildeWn}
\end{equation}
where $L_n$ are the Laguerre polynomials.
So, whatever result we obtain for $W_n$, replacing the parameter $a$ 
by $1/2$ should yield~\eqref{tildeWn}.

\section{Wigner distribution function for the ground state and the first excited state}
In this section, we compute an explicit expression for $W_0(p,q)$.
This will already show the techniques involved in the computation of 
the Wigner d.f.~for arbitrary states $|n\rangle$,
but the resulting calculations are not quite so lengthy.  To reiterate,
the goal is to compute
\begin{equation}
	W_0(p,q) = \frac{1}{4\pi^2} \iint \langle 0 | \mexp{i\X} | 0 \rangle
	\mexp{-i(\lambda p + \mu q)}\, d\lambda d\mu.
	\label{W_0}
\end{equation}
Thus, obviously, in a first step we are going to determine the 
matrix elements $\langle 0 | \exp(i\X) | 0 \rangle$, and in a second step
the integration will be performed.

The operator $\exp(i\X)$ being defined as 
\begin{equation*}
\mexp{i\X} = \sum_{k\geq 0} \frac{i^k \X^k}{k!} 
\end{equation*}
we first try to find $\langle 0 | \X^k | 0\rangle$.  To this end, we write 
$\X = \lambda \p + \mu \q$ in terms of $\b^\pm$:
\begin{equation}
\X = \alpha^+ \b^+ + \alpha^- \b^-\ \text{with}\ 
\alpha^{\pm} = \frac{\mu\pm i\lambda}{\sqrt{2}}.
	\label{Xb}
\end{equation}

It is then immediately clear that 
\begin{equation}
	\langle 0 | \X^{2k+1} | 0\rangle = 0
	\label{0-X-odd}
\end{equation}
since an odd number of applications of $\b^+$ and $\b^-$ on $|0\rangle$ 
does not yield a term corresponding to $|0\rangle$ in the resulting expansion,
as can be seen from~\eqref{bpm-actions}.
On the other hand, using~\eqref{bpm-actions}, one finds easily that
\begin{equation*}
\begin{split}
\langle 0 | \X^0 | 0\rangle & =  1, \\
\langle 0 | \X^2 | 0\rangle & =  2\alpha^+\alpha^- a, \\
\langle 0 | \X^4 | 0 \rangle & =  (2\alpha^+\alpha^-)^2 a(a+1), \\
\langle 0 | \X^6 | 0 \rangle & =  (2\alpha^+\alpha^-)^3 a(a+1)(a+2).
\end{split}
\end{equation*}
This suggests, as will be proved later, that one has
\begin{equation}
	\langle 0 | \X^{2k} | 0 \rangle  =(2\alpha^+\alpha^-)^k\, (a)_k = 
	(\lambda^2+\mu^2)^k\, (a)_k.
	\label{0-X-even}
\end{equation}

Using~\eqref{0-X-odd} and~\eqref{0-X-even}, one thus finds that
\begin{equation}
\begin{split}
\langle 0 | \mexp{i\X} | 0 \rangle & = 
\langle 0 | \sum_{k\geq 0} \frac{i^k \X^k}{k!} | 0 \rangle 
= \sum_{k\geq 0} \frac{(-1)^k}{(2k)!} \langle 0 | \X^{2k} | 0 \rangle  =
 \sum_{k\geq 0} \frac{(-1)^k(\lambda^2+\mu^2)^k\, (a)_k}{(2k)!}\\
 & = 
\sum_{k\geq0} \frac{(a)_k}{k!(1/2)_k} (-\frac{\lambda^2+\mu^2}{4})^k = 
{}_1F_1\left(\atop{a}{1/2};  -\frac{\lambda^2+\mu^2}{4}\right).
	\label{0-exp(iX)}
\end{split}
\end{equation}
where we have used the fact that $(2k)! = (1/2)_k \, k!\, 4^{k}$.
This means that we now have:
\begin{equation*}
W_0(p,q) = \frac{1}{4\pi^2} \iint 
{}_1F_1\left(\atop{a}{1/2};  -\frac{\lambda^2+\mu^2}{4}\right)
\mexp{-i(\lambda p + \mu q)}\, d\lambda d\mu.
\end{equation*}

In order to compute this integral, it is useful to introduce a quadratic decaying
exponential into the integrand by applying Kummer's transformation~\eqref{Kummer}
on the confluent hypergeometric series.  One then has:
\begin{equation*}
\begin{split}
W_0(p,q) & = \frac{1}{4\pi^2} \iint 
{}_1F_1\left(\atop{1/2-a}{1/2};  \frac{\lambda^2+\mu^2}{4}\right)
\mexp{-(\lambda^2+\mu^2)/{4}}
\mexp{-i(\lambda p + \mu q)}\, d\lambda d\mu \\
& = \frac{1}{4\pi^2} \sum_{k\geq0}  \frac{(1/2-a)_k}{k!(1/2)_k 4^k}
\iint \mexp{-(\lambda^2+\mu^2)/{4}}
\mexp{-i(\lambda p + \mu q)} (\lambda^2+\mu^2)^k\, d\lambda d\mu, 
\end{split}
\end{equation*}
where we have swapped summation and integration.  We assume that the parameter
$a$ is such that this is possible.  In any case, when the parameter $a$ is
of the form
\begin{equation}
	a = 1/2+m\ \text{with}\ m\in\Z_+,
\label{a}
\end{equation}
this can certainly 
be done since one is then only dealing with finite summations.
In Appendix A, it is shown that
\begin{equation}
	\frac{1}{4\pi^2}\iint e^{-\fracs{(\lambda^2+\mu^2)}{4}}
	e^{-i(\lambda p + \mu q)} (\lambda^2+\mu^2)^k\, d\lambda d\mu = 
	\frac{1}{\pi}e^{-p^2-q^2} k!\, 4^k\, L_k(p^2+q^2),  
	\label{integral}
\end{equation}
such that 
\begin{equation}
	W_0(p,q) = \frac{1}{\pi}e^{-p^2-q^2} 
\sum_{k\geq 0} \frac{(1/2-a)_k}{(1/2)_k} L_k(p^2+q^2).
	\label{W0-res}
\end{equation}

It is important to remark that, apart from the exponential factor, the
expression~\eqref{W0-res} is a polynomial in $p^2+q^2$ whenever 
the parameter $a$ is of the form~\eqref{a}. 
In this case, it is possible to simplify the expression~\eqref{W0-res}
even further.
Expand the Laguerre polynomials as  single sums and switch the two (finite) summations.
On the resulting inner sum Vandermonde's summation theorem~\eqref{Vander} 
can be applied.  After some
further easy manipulations one then finds:
\begin{equation}
W_0(p,q) = \frac{1}{\pi} e^{-p^2-q^2} \frac{1}{1-2m}\,
{}_1F_1(\atop{-m}{3/2-m}; p^2+q^2),
\quad\text{with}\ a = 1/2+m.
\label{W0-m}
\end{equation}
Clearly, in the canonical case (i.e.~$m=0$) the resulting expression reduces correctly to $\tilde W_0(p,q)$.

To compute $W_1(p,q)$, we first have to find an expression for the 
matrix elements $\langle 1 | \X^k | 1\rangle$.  Again, one finds  
the matrix elements with odd powers of $\X$ to be zero.  For the even powers,
one finds using~\eqref{bpm-actions} and~\eqref{Xb}:
\begin{equation*}
\begin{split}
\langle 1 | \X^0 | 1\rangle & =  1, \\
\langle 1 | \X^2 | 1\rangle & =  2\alpha^+\alpha^- (a+1), \\
\langle 1 | \X^4 | 1 \rangle & =  (2\alpha^+\alpha^-)^2 (a+1)(a+2), \\
\langle 1 | \X^6 | 1 \rangle & =  (2\alpha^+\alpha^-)^3 (a+1)(a+2)(a+3),
\end{split}
\end{equation*}
indicating that
\begin{equation*}
	\langle 1 | \X^k | 1\rangle = \langle 0 | \X^k | 0\rangle|_{a\to a+1}.
\end{equation*}
Since these matrix elements are the only place where the parameter $a$ appears, 
the expression for $W_1(p,q)$ is also found by replacing $a$ with $a+1$ in the
expression for $W_0(p,q)$:
\begin{equation}
W_1(p,q) = W_0(p,q)|_{a\to a+1}.
	\label{W1}
\end{equation}

\section{Wigner distribution function for arbitrary states $|n\rangle$}
We now would like to find an explicit expression for the following integral:
\begin{equation*}
	W_n(p,q) = \frac{1}{4\pi^2} \iint \langle n | \mexp{i\X} | n \rangle
	\mexp{-i(\lambda p + \mu q)} d\lambda d\mu.
\end{equation*}
As is clear from the previous section, the key in calculating this integral
is finding an expression for the matrix elements $\langle n | \X^k | n \rangle$.
As before, matrix elements with odd powers of $\X$ are easily seen to be zero.
It will also be the case that matrix elements for a state $|2n+1\rangle$ will
be determined from those with state $|2n\rangle$ simply by replacing $a$ 
with $a+1$.  We will now give two different expressions for the
matrix elements $\langle 2n | \X^{2k} | 2n \rangle$.  We will show that 
they are equivalent and will compute the Wigner d.f.~using
each of them.

The proof that the proposed expressions are in fact the correct matrix elements 
is relegated to an appendix, despite the importance of the result to 
the calculations.  The proof, although in itself not difficult, requires
rather lengthy calculations as one also needs the off-diagonal matrix elements.
Basically, it involves checking that a certain expression satisfies the
recurrence relation for the matrix elements (as well as the correct boundary 
conditions).

We have the following expressions for the matrix elements 
$\langle 2n | \X^{2k} | 2n \rangle$:
\begin{equation}
\langle 2n | \X^{2k} | 2n \rangle = (\lambda^2+\mu^2)^k 
\sum_{\atop{j=0}{2j\leq k}}^n 
\frac{(2j)!}{j!} \binom{n}{j}\binom{k}{2j} (a+n-j)_j (a+2n)_{k-2j}, 
	\label{J}
\end{equation}
or alternatively
\begin{equation}
\langle 2n | \X^{2k} | 2n \rangle = (\lambda^2+\mu^2)^k 
\sum_{j=0}^{\min(n,k)} \frac{1}{j!} \binom{n}{j} (a+j)_{k-j} (k+1-j)_{2j}.
	\label{S}
\end{equation}
Note that both expressions reduce to~\eqref{0-X-even} for $n=0$.  To see that 
both expressions are indeed equivalent, it is easiest to write them in 
hypergeometric notation.  For formula~\eqref{S} we have
\begin{equation*}
\langle 2n | \X^{2k} | 2n \rangle = (\lambda^2+\mu^2)^k\,
(a)_k\,\,\, {}_3F_2\left(\atop{-k,k+1, -n }{1,a};1 \right),
\end{equation*}
whilst formula~\eqref{J} is
\begin{equation*}
\langle 2n | \X^{2k} | 2n \rangle = (\lambda^2+\mu^2)^k\,
(a+2n)_k\,\,\,
{}_4F_3\left(\atop{-n,-k/2,1/2-k/2,1-n-a}{1,1/2-n-a/2-k/2,1-n-a/2-k/2 };1 \right).
\end{equation*}
So, we would like a transformation for hypergeometric series that relates a series
of type ${}_3F_2$ with one of type ${}_4F_3$.  In Krattenthaler's list~\cite{HYP},
transformation T3231 does just that and is indeed applicable to the 
${}_3F_2$-series at hand.  It will turn out that we will need to apply yet 
another transformation, namely T4301 from the list, to the resulting series.
Finally, a doubling formula for rising factorials, see e.g.~\cite[I.25]{Slater},
has to be applied.
We thus have:
\begin{equation*}
\begin{split}
&{}_3F_2\left(\atop{-k,k+1, -n }{1,a};1 \right)  = 
{}_4F_3\left(\atop{-n,-k/2, k/2+1/2, a+n}{1,a/2,a/2+1/2}  ;1\right) \\
&\quad = 
\frac{(a/2+k/2, a/2+k/2+1/2)_n}{(a/2,a/2+1/2)_n}
\,\,{}_4F_3\left( \atop{-n,-k/2,1/2-k/2,1-n-a}{1,1/2-n-a/2-k/2,1-n-a/2-k/2 }  ;1\right) \\
&\quad = \frac{(a+k)_{2n}}{(a)_{2n}} 
{}_4F_3\left( \atop{-n,-k/2,1/2-k/2,1-n-a}{1,1/2-n-a/2-k/2,1-n-a/2-k/2 }  ;1\right).
\end{split}
\end{equation*}
This shows that formulas~\eqref{J} and~\eqref{S} are indeed equivalent.

The next step is to compute expressions for $\langle 2n | \exp(i\X) | 2n\rangle$,
by using the series expansion for the operator $\exp(i\X)$.  We first 
use~\eqref{J}:
\begin{equation}
\begin{split}
\langle 2n | \mexp{i\X} | 2n\rangle &  = 
\sum_{k\geq0} \frac{(-1)^k}{(2k)!} \langle 2n | \X^{2k} | 2n\rangle \\
&= \sum_{j=0}^n \binom{n}{j} \frac{(2j)!(a+n-j)_j}{(4j)!j!}(\lambda^2+\mu^2)^{2j}
\, {}_1F_1\left( \atop{a+2n}{2j+1/2}; -\frac{\lambda^2+\mu^2}{4}\right).
	\label{A28}
\end{split}
\end{equation}
This simply follows from swapping the summation order and writing the inner 
sum again as a hypergeometric series.  The same procedure is applicable 
to formula~\eqref{S},  but here one gets:
\begin{equation}
	\langle 2n | \mexp{i\X} | 2n\rangle = 
\sum_{j=0}^n \binom{n}{j}\frac{(-1)^j}{j!}(\lambda^2+\mu^2)^{j}
\, {}_2F_2\left( \atop{a+j, 1+2j}{j+1/2, 1+j}; -\frac{\lambda^2+\mu^2}{4}\right).
	\label{A27}
\end{equation}

Basically, all that is left to do is to integrate these expressions against the
complex exponential $\exp(-i(\lambda p+\mu q))$.  For the
expression~\eqref{A28}, we proceed as in the case of the ground state.  We
apply Kummer's transformation~\eqref{Kummer} to each of the confluent
hypergeometric series.  One will then get a double summation over integrals of
the type~\eqref{integral}. (Here, we thus again assume that the order of
summation and integration may be switched, which is certainly the case when the
parameter $a$ is of the form~\eqref{a}.) Application
of~\eqref{integral} for each of the integrals in the series, will yield a
triple sum expression for $W_{2n}(p,q)$, with one summation \lq\lq hidden\rq\rq\
in the Laguerre polynomial.  Explicitly, we have the following
result:
\begin{equation}
W_{2n}(p,q) = \frac{1}{\pi} e^{-p^2-q^2} 
\sum_{j=0}^n \binom{n}{j}\frac{(a+n-j)_j}{j!(1/2-a-2n)_{2j}} 
\sum_{k\geq 2j} \frac{k!\,(1/2-a-2n)_k}{(1/2)_k(k-2j)!} L_k(p^2+q^2).
	\label{A29}
\end{equation}

As a sanity check, we would like to see that for $a=1/2$ formula~\eqref{A29} 
reduces to $\tilde W_{2n}(p,q)$.  This means that the summation part of~\eqref{A29}
should equal $L_{2n}(2p^2+2q^2)$.  This is not immediately clear, so we substitute
$a=1/2$ and write the Laguerre polynomial as an explicit summation and the triple
sum then becomes:
\begin{equation*}
	\sum_{k,l} \frac{(-2n)_k(-k)_l (p^2+q^2)^l}{(1/2)_k l!^2} 
	{}_2F_1\left(\atop{-k/2,-k/2+1/2}{1} ;1\right),
\end{equation*}
where the summation over $j$ is now performed first. 
The Gauss hypergeometric series, which terminates because either $-k/2$ or
$-k/2+1/2$ is a non positive integer, can be summed in a unified way  by 
applying transformation~\cite[T2140]{HYP} first:
\begin{equation*}
	{}_2F_1\left(\atop{-k/2,-k/2+1/2}{1} ;1\right) = 2^{-k} 
	{}_2F_1\left(\atop{-k,-k}{1} ;1\right) = 2^{-k} \frac{(1+k)_k}{k!}.
\end{equation*}
Apply this summation, and one gets a double summation.  The summation 
over the index $k$ can then be summed using the binomial theorem, and 
the result follows, showing that for $a=1/2$ expression~\eqref{A29} reduces
correctly to the known expression~\eqref{tildeWn}.

Can we also do the integration starting from~\eqref{A27}?  Clearly, we need
a Kummer-like transformation for ${}_2F_2$-series that introduces a quadratic
decaying exponential.
Fortunately, such a transformation was published quite recently by 
Paris~\cite{paris}.  His main transformation expresses one ${}_2F_2$-series
as an exponential function multiplied by an infinite series containing
${}_2F_2$-series.  We, however, only need a simpler version where a finite 
sum of ${}_2F_2$-series is involved:
\begin{equation}
{}_2F_2\left(\atop{a,c+j}{b,c} ;z\right) = 
e^z \sum_{k=0}^j \binom{j}{k} \frac{z^k}{(c)_k}
\, {}_2F_2\left(\atop{b-a,c+j}{b,c+k} ;-z\right),\quad\text{with}\ j\in\Z_+.
	\label{Paris}
\end{equation}

As is stands, expression~\eqref{A27} is a double sum.  After applying~\eqref{Paris},
it will be a triple sum, and since integrating using~\eqref{integral} introduces 
an extra summation, the resulting expression will be a four-fold summation.
Since this computation is quite similar to the ones we have already done,
we  just state the result:
\begin{equation*}
W_{2n}(p,q) = 
\frac{1}{\pi}e^{-p^2-q^2} \sum_{j,l} \frac{(-n)_j(1/2-a)_l(j+1)_{j+l}}{j!l!(1/2)_{j+l}}
\sum_{k=0}^j (-1)^k\binom{j}{k} L_{j+l+k}(p^2+q^2).
\end{equation*}
There now exist quite a number of formulas relating sums of Laguerre polynomials 
to another Laguerre polynomial, one of them being the following~\cite[4.4.1.5]{Prud-Vol2}:
\begin{equation*}
\sum_{k=0}^n (-1)^k \binom{n}{k} L_{r+k}^{(\alpha)}(z) = (-1)^n L_{r+n}^{(\alpha-n)}(z).
\end{equation*}
This identity is also valid for $\alpha = 0$, provided one defines the 
Laguerre polynomials as in~\eqref{Lag-gen}.
Application of this identity then leads immediately to:
\begin{equation}
W_{2n}(p,q) = \frac{1}{\pi}e^{-p^2-q^2} 
\sum_{j=0}^n \binom{n}{j} \frac{(j+1)_j}{(1/2)_j}\sum_{k\geq 0} 
\frac{(1/2-a)_{k}(2j+1)_{k}}{k!\, (1/2+j)_{k}} L_{k+2j}^{(-j)}(p^2+q^2). 
	\label{A31}
\end{equation}
Also in this case, reduction to the known result for $a=1/2$ can be verified, and is 
easier than in the previous case, since the identity needed is 
recorded as~\cite[4.4.1.16]{Prud-Vol2}.

We thus have found closed expressions for $W_{2n}(p,q)$ namely~\eqref{A29} 
and~\eqref{A31}.  {}From this we now 
immediately find the corresponding expressions $W_{2n+1}(p,q)$.  We state
the following:
\begin{equation}
W_{2n+1}(p,q) = W_{2n}(p,q)|_{a\to a+1}.
	\label{W2n+1}
\end{equation}
This follows immediately from the fact, shown in Appendix B, that
\begin{equation}
\langle 2n+1 | \X^{k} | 2n+1\rangle =\langle 2n | \X^{k} | 2n\rangle|_{a\to a+1},
\end{equation}
together with the observation that this is the only place in which the
parameter $a$ plays a role in the computation of the Wigner distribution 
function.


\section{Discussion and conclusions}

Parastatistics is one of the alternative approaches in particle statistics
proposing that there are particles obeying statistics different from
Bose-Einstein or Fermi-Dirac~\cite{ohnuki}.  It involves an additional
parameter (often referred to as the order of statistics), and for one
particular value of this parameter one recovers the common Bose-Einstein or
Fermi-Dirac statistics.  In the current case, where we just considered one
parabose oscillator, this parameter is $a$, and for $a=1/2$ one is reduced to
the canonical quantum oscillator.

The simple case treated here can also be viewed as a deviation from canonical
quantum mechanics.  The one-dimensional quantum oscillator is treated as a
Wigner Quantum System, and allows several representations of $\osp(1|2)$
labelled by a positive number~$a$.  In some sense, $a$ (or $a-1/2$) can be
viewed as a quantity measuring the deviation from canonical quantum mechanics,
see~\eqref{action-comm-pq}. 

We have now obtained, for the first time, explicit expressions of the Wigner
d.f.\ for the parabose oscillator stationary states.  This allows one to
consider ``phase space densities'' for a ``parastatistics quantum system'' or
for a ``non-canonical quantum system''.

Let us now examine some plots of the Wigner distribution functions.
First of all, note that the Wigner d.f.\ $W_n(p,q)$ is, just as in the canonical case,
a function of $p^2+q^2$ only. 
So the 3D-plots of $W_n(p,q)$ in $(p,q)$-space are rotational invariant. 
This is why it is more convenient to give only two-dimensional plots of
$W_n(p,q)$ as a function of $r$, where
\[
r = \sqrt{p^2+q^2}.
\]

In Figure~1 we have plotted the ground state Wigner d.f.\ $W_0(p,q)$ of the
one-dimensional parabose oscillator for values of the parameter $a$ equal to
$1/2,\;3/2,\;5/2$ and $7/2$ (or $m=0,1,2,3$).  One can see that for $a=1/2$
($m=0$, the well-known canonical case) one finds the usual Gaussian behaviour
of the distribution function.  As $a$ (or $m$) increases, the shape changes
quite significantly.  For values of $a$ larger than $1/2$, the shape
of $W_0(p,q)$ is rather similar to the shape of the Wigner d.f.\ for the first
excited state of the canonical harmonic oscillator.  
In fact, it follows from~\eqref{W1} that $W_0(p,q)$ for $a=3/2$ is equal to $W_1(p,q)$
for $a=1/2$, so in that specific case it is identical to the canonical Wigner d.f.\
of the first excited state.
As $a$ increases, the
maximum value of $W_0(p,q)$ moves further away from the origin.  The ground
state of the parabose oscillator behaves as the first excited state of a
canonical quantum oscillator but with increasing energy as $a$ increases.

In Figure~2 we plot the distribution functions $W_n(p,q)$ for a fixed $a$-value
(namely $a=3/2$), and for a number of stationary states $(n=0,1,2,3$).  Note
the separation of even and odd states near the origin. 

The properties of the Wigner distribution functions are in agreement with the 
``wave-mechanical representation'' for the parabose oscillator, as discussed 
in~\cite[Chapter~23]{ohnuki}.
The wave functions in the position representations have also been given here
in~\eqref{wave}.
Note that these wave functions are also the solutions of a wave equation of the form
\begin{equation}
\left( -\frac{1}{2} \frac{d^2}{dq^2} + \frac{1}{2} q^2 + \frac{1}{2}\frac{g(g-1)}{q^2} \right)
\Psi(q) = E \Psi(q).
\label{wave-eq}
\end{equation}
The even solutions are the functions $\Psi_{2n}^{(a)}(q)$ with $a=g+1/2$ and $E=g+1/2+2n$, and
the odd solutions are the functions $\Psi_{2n+1}^{(a)}(q)$ with $a=g-1/2$ and $E=g+1/2+2n$.
Equation~\eqref{wave-eq} is often referred to as that of the non-relativistic
singular harmonic oscillator, or of the 
 Calogero-Sutherland model~\cite{calogero69, calogero71, sutherland}.
The singular behaviour of the last term of the Hamiltonian in~\eqref{wave-eq} is once more
in agreement with the negative values of the Wigner distribution function near the origin.

To summarize, we have in this paper constructed the Wigner distribution function for
the one-dimensional parabose oscillator.
This is a quantum system for which the Hamiltonian is given by
the standard expression~\eqref{H}, and for which the position and momentum operators
$\hat p$ and $\hat q$ do not satisfy the canonical commutation relations, but
instead the more general compatibility conditions~\eqref{CCs}.
This non-canonical situation makes the definition (and computation) of the Wigner
function more difficult.
The approach of Section~3 allowed us to define the right Wigner function, satisfying
all necessary properties.
The computation of the Wigner function $W_n(p,q)$ for stationary states is still
quite involved, and uses some techniques from hypergeometric series.
The main results of the computation are given in~\eqref{A29}, \eqref{A31} and~\eqref{W2n+1}.
On the basis of some plots, the behaviour of the Wigner distribution function is 
further discussed in the context of parastatistics or in the context of the
wave-mechanical representation of the parabose oscillator.

\begin{figure}
\includegraphics[width=0.90\textwidth]{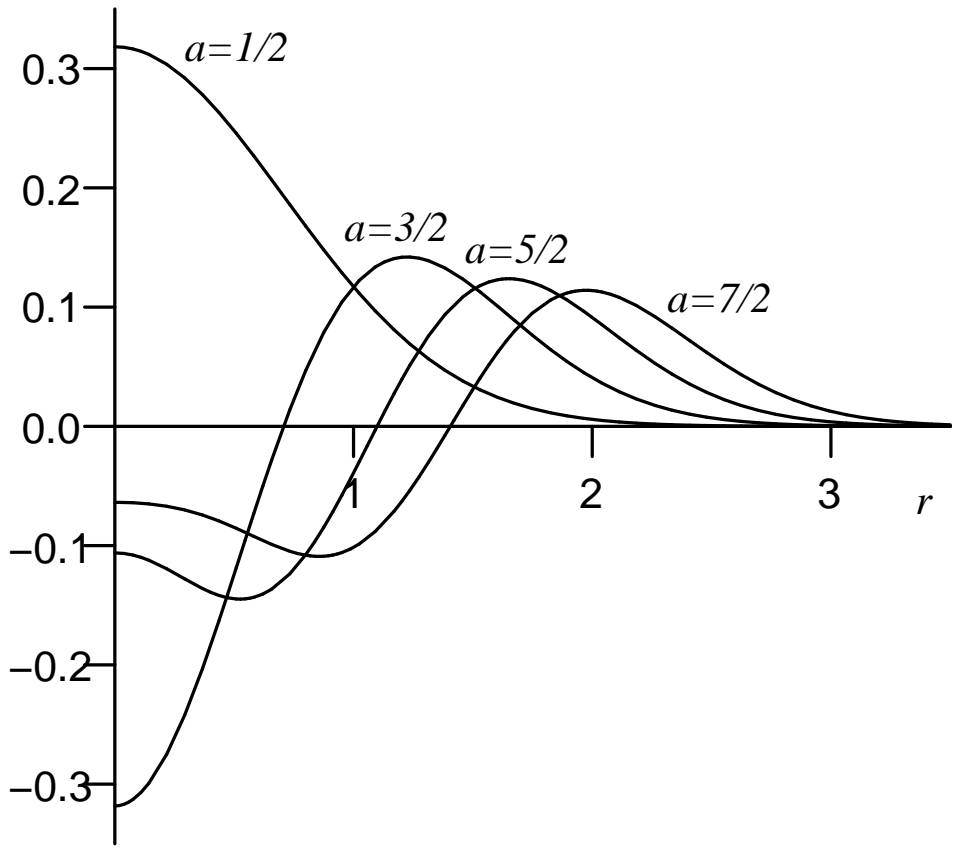}
\caption{The ground state Wigner function of the one-dimensional parabose oscillator, for values of $a=1/2,\;3/2,\;5/2,\;7/2$ ($m=0,\;1,\;2,\;3$).}
\end{figure}

\begin{figure}
\includegraphics[width=0.90\textwidth]{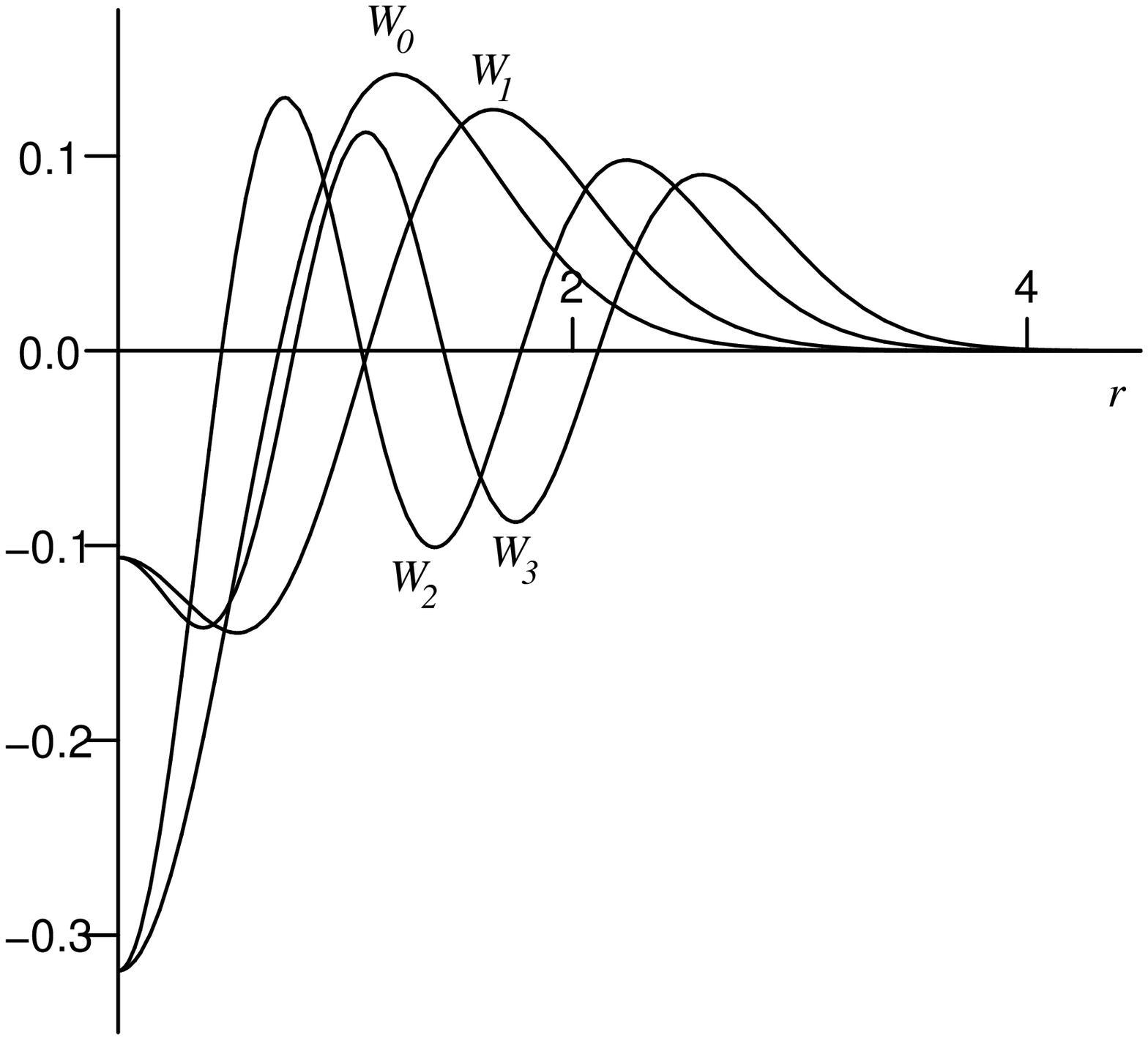}
\caption{The Wigner function of the stationary states of the one-dimensional
parabose oscillator $W_n(p,q)$, for fixed value of $a=3/2$ ($m=1$) and $n=0,1,2,3$. }
\end{figure}


\section*{Acknowledgments}

One of the authors (E.J.) would like to acknowledge that this work is
performed in the framework of the Fellowship 05-113-5406 under the
INTAS-Azerbaijan YS Collaborative Call 2005.  Another author (S.L.) was supported
by project P6/02 of the Interuniversity Attraction Poles Programme 
(Belgian State -- Belgian Science Policy).

\section*{Appendix A}
We have to show that
\begin{equation}
	\frac{1}{4\pi^2}\iint e^{-\fracs{(\lambda^2+\mu^2)}{4}}
	e^{-i(\lambda p + \mu q)} (\lambda^2+\mu^2)^k\, d\lambda d\mu = 
	\frac{1}{\pi}e^{-p^2-q^2} k!\, 4^k\, L_k(p^2+q^2),  
	\label{integral-bis}
\end{equation}
for $k\in\Z_+$ and $p$ and $q$ real.  Let us call the left hand side
of the expression~\eqref{integral-bis} $I$, and apply the binomial
theorem to the factor $(\lambda^2+\mu^2)^k$:
\begin{equation*}
I = \sum_{j=0}^k \binom{k}{j} 
\left(\frac{1}{2\pi} \int e^{-\lambda^2/4 - ip\lambda} \lambda^{2j}\,d\lambda\,\right)
\left(\frac{1}{2\pi} \int e^{-\mu^2/4 - iq\mu} \mu^{2k-2j}\,d\mu\right).
\end{equation*}
Each of the two simple integrals above has a known expression in terms of Hermite 
polynomials~\cite[2.3.15.10]{Prud-Vol1}:
\begin{equation*}
\frac{1}{2\pi} \int e^{-ax^2 - ibx} x^{2n}\, dx = 
\frac{(-1)^n e^{-b^2/4a}}{\sqrt{\pi} 2^{2n+1} a^n \sqrt{a}}\,
H_{2n}(-\frac{b}{2\sqrt{a}}),
\quad a > 0,\, b\in\Real
\end{equation*}
yielding
\begin{equation*}
I = \frac{(-1)^k}{\pi}e^{-p^2-q^2} \sum_{j=0}^k \binom{k}{j} H_{2j}(-p) H_{2k-2j}(-q).
\end{equation*}
Finally, there is a known summation formula for Hermite polynomials~\cite[4.5.2.4]{Prud-Vol1}:
\begin{equation*}
\sum_{j=0}^k \binom{k}{j} H_{2j}(x) H_{2k-2j}(y) = (-1)^k k! 4^k L_k(x^2+y^2).
\end{equation*}
Application of this summation formula immediately yields the desired 
result~\eqref{integral-bis}.

\section*{Appendix B}
In this Appendix, it is our aim to prove formula~\eqref{S}.  We will do so by 
given formulas for the matrix elements $\langle 2n+2l | \X^{2k} | 2n\rangle$, 
with $l\in \Z$, which will be seen to reduce to formula~\eqref{S} when $l = 0$.
The proof relies on the fact that the formulas given below satisfy the
same recurrence and the same boundary conditions as the matrix elements themselves.

With $\X$ as before, one sees that 
\begin{equation*}
\X^2 = (\alpha^+)^2(\b^+)^2 + \alpha^+\alpha^-\{\b^+,\b^-\} + (\alpha^-)^2(\b^-)^2,
\end{equation*}
and hence one gets, using~\eqref{bpm-actions} and~\eqref{action-anticomm}:
\begin{equation*}
\begin{split}
\langle 2n+2l | \X^{2k} | 2n\rangle & = \langle 2n+2l | \X^{2k-2} \X^2 | 2n\rangle \\
& = (\alpha^+)^2\langle 2n+2l | \X^{2k-2}(\b^+)^2  | 2n\rangle +
\alpha^+\alpha^-\langle 2n+2l | \X^{2k-2} \{\b^+,\b^-\}  | 2n\rangle \\ 
& \qquad + (\alpha^-)^2\langle 2n+2l | \X^{2k-2}(\b^-)^2  | 2n\rangle  \\
& = 2(\alpha^+)^2\sqrt{(n+a)(n+1)}\, \langle 2n+2l | \X^{2k-2}  | 2n+2\rangle \\
&\qquad+
2\alpha^+\alpha^- (2n+a) \, \langle 2n+2l | \X^{2k-2} | 2n\rangle \\ 
& \qquad\qquad + 2(\alpha^-)^2 \sqrt{n(n+a-1)}\,  \langle 2n+2l | \X^{2k-2}  | 2n-2\rangle. 
\end{split}
\end{equation*}
So, if we now denote $\langle 2n+2l | \X^{2k} | 2n\rangle$ as $F_{k,l}(n)$, the
recurrence that $F_{k,l}(n)$  satisfies is:
\begin{equation}
\begin{split}
	F_{k,l}(n) & = 2(\alpha^+)^2\sqrt{(n+a)(n+1)}\, F_{k-1,l-1}({n+1}) +
2\alpha^+\alpha^- (2n+a) \, F_{k-1,l}(n)  \\
&\qquad + 2(\alpha^-)^2 \sqrt{n(n+a-1)}\,  F_{k-1,l+1}({n-1}).
	\label{rec-F}
\end{split}
\end{equation}
{}From the fact that we are dealing with orthonormal states $|n\rangle$, it 
follows immediately that 
\begin{equation}
F_{0,0}(n) = 1,
\label{bound1}
\end{equation}
and also from the actions~\eqref{bpm-actions} and~\eqref{action-anticomm} it
is clear that
\begin{equation}
F_{k,l}(n) = 0,\quad\text{if}\ |l| > k.
	\label{bound2}
\end{equation}
These two boundary conditions determine the unique solution of the 
recurrence~\eqref{rec-F}.  We will now give explicit formulas that
satisfy the boundary conditions~\eqref{bound1} and~\eqref{bound2}, 
and we will verify that these formulas also satisfy~\eqref{rec-F}. 
The verification of that last fact will boil down to checking that
two polynomials are equal.  As a polynomial basis, we will however
not choose the classical basis $x^k$, but rather we will work
with the polynomials $(x)_k$.

We now state the formulas,
\begin{equation}
\begin{split}
\langle 2n+2l | \X^{2k} | 2n\rangle & = 
(-1)^l 2^k (\alpha^+)^{k+l}(\alpha^-)^{k-l} \sqrt{(n+1)_l(n+a)_l} \\
& \quad\times
\sum_{j=0}^{\min(n,k-l)} \frac{(a+l+j)_{k-l-j}(-n)_j(-k)_{l+j}(k+l+1)_j}{(l+j)!j!}
\label{J1a}
\end{split}
\end{equation}
and
\begin{equation}
\begin{split}
\langle 2n-2l | \X^{2k} | 2n\rangle & = 
(-1)^l 2^k (\alpha^+)^{k-l}(\alpha^-)^{k+l} \sqrt{(n-l+1)_l(n-l+a)_l} \\
& \quad\times
\sum_{j=0}^{\min(n-l,k-l)} \frac{(a+l+j)_{k-l-j}(-n+l)_j(-k)_{l+j}(k+l+1)_j}{(l+j)!j!},
\label{J1b}
\end{split}
\end{equation}
where, in both cases, $l\geq0$.
It is immediately clear that formulas~\eqref{J1a} and~\eqref{J1b} satisfy 
the boundary conditions~\eqref{bound1} and~\eqref{bound2}.  Also, the reduction
to~\eqref{S} for $l=0$ is readily seen.

Since the explicit formulas for the matrix elements are different depending
on $l$ being positive or not, checking that~\eqref{J1a} and~\eqref{J1b} indeed
satisfy~\eqref{rec-F} will involve three cases: $l >0$, $l = 0$ and $l < 0$.
Since all three are very similar, we will only deal with the case $l > 0$ here.
We will abbreviate the summation in~\eqref{J1a} as $S_{k,l}(n)$, then substituting
the expression~\eqref{J1a} in~\eqref{rec-F} allows simplification of the 
powers of $2$ and $\alpha^\pm$ as well as the square roots, and leads
to the following that needs to be checked:
\begin{equation*}
S_{k,l}(n) = -S_{k-1,l-1}({n+1})+(2n+a)S_{k-1,l}(n)-n(n+a-1)S_{k-1,l+1}({n-1}).
\end{equation*}
This is a polynomial identity in the variable $-n$. Let us denote $-n$ as $x$.
The coefficients of the polynomials $(x)_j$ are immediately clear for 
the lhs.  For the right hand side, one can use the following easy identities:
\begin{equation*}
(x-1)_j = (x)_j - j(x)_{j-1},\quad\text{and}\quad
x(x)_j = (x)_{j+1} - j(x)_j.
\end{equation*}
This allows to extract the coefficient of $(x)_j$ in the right hand side.  
This coefficient is a rational function in the variables $a$, $k$, $j$ and $l$.
After clearing the denominator and simplifying the remaining rising factorials,
one ends up with a polynomial of a low degree that needs to be identically zero,
and that is indeed the case.

So, although the checking is rather tedious, it is at the same time also simple
and indeed shows that our formulas for the matrix elements are correct.

We end this appendix by noting that one indeed has:
\begin{equation*}
\langle 2n+1+2l | \X^{k} | 2n+1\rangle =\langle 2n+2l | \X^{k} | 2n\rangle|_{a\to a+1}.
\end{equation*}
Indeed, it is easily checked that the recurrence satisfied by 
$G_{k,l}(n) = \langle 2n+1+2l | \X^{k} | 2n+1\rangle$ is identical
to~\eqref{rec-F} but with $a$ replaced by $a+1$.  Also, the boundary 
conditions for $G$ are identical to~\eqref{bound1} and~\eqref{bound2}
and as these do not involve $a$ the statement is now proved.

\section*{Appendix C}

In this Appendix, we are briefly going to discuss some aspects of the convergence 
of the main expressions~\eqref{W0-res} and~\eqref{A29}. 
The main result is that expressions~\eqref{W0-res} and~\eqref{A29}
converge absolutely when $a>1$ for all $p$ and $q$.  In the origin, i.e.~$(p,q) = (0,0)$,
the condition $a>1$ is also necessary for convergence.  For the odd
states, we thus have absolute convergence for the expressions
derived from~\eqref{W0-res} and~\eqref{A29} everywhere for $a>0$.
The only remaining question is thus what happens for 
the even states when $(p,q) \neq (0,0)$ in the case $0<a<1$ (and $a\neq 1/2$).

First, let us 
recall the well known fact that if one has two series of positive terms
$\sum_{k} b_k$ and $\sum_{k} c_k$ such that $b_k \leq c_k$, then 
convergence of $\sum_{k} c_k$ implies that of $\sum_{k} b_k$.  Secondly, there
are some well known tests for convergence of positive series, one of them 
is Raabe's test which can be used if the ratio test fails:
let $\sum_{k} b_k$ be a series of positive terms and let:
\begin{equation*}
\lim_{k\to\infty} k( \frac{b_k}{b_{k+1}} - 1) = L,
\end{equation*}
one then has that $\sum_{k}b_k$ converges if $L > 1$ and diverges whenever $L < 1$.
When $L=1$, the test is inconclusive.

Let us abbreviate $p^2+q^2$ as $t$.  The convergence of~\eqref{W0-res} then 
reduces to the convergence of 
\begin{equation}
\sum_{k\geq 0} \frac{(1/2-a)_k}{(1/2)_k} L_k(t) = \sum_{k\geq 0} b_k.
	\label{W0-conv}
\end{equation}
Next, we are going to use an inequality for the Laguerre polynomials $L_k(t)$ 
which gives an upper bound for $|L_k(t)|$ not depending on $k$~\cite[10.18(3)]{Bateman}:
\begin{equation}
|L_k(t)| \leq e^{t/2},\ \text{for}\ t\geq 0.
	\label{Lag-ineq}
\end{equation}
This gives us:
\begin{equation*}
	|b_k| = \left|\frac{(1/2-a)_k}{(1/2)_k} L_k(t)\right| \leq 
	\left|\frac{(1/2-a)_k}{(1/2)_k}\right| e^{t/2} = c_k.
\end{equation*}
The ratio test fails on the series $\sum_k c_k$, but Raabe's test
gives:
\begin{equation*}
\lim_{k\to\infty} k( \frac{c_k}{c_{k+1}} - 1) = 
\lim_{k\to\infty} k\left( \frac{|1+2k|}{|1-2a+2k|} - 1 \right)
 = \lim_{k\to\infty} k\left( \frac{1+2k}{1-2a+2k} - 1 \right) = a.
\end{equation*}
So clearly, the series $\sum_k c_k$ converges if $a>1$, and hence the
series~\eqref{W0-conv} is absolutely convergent in that case.
Note that the fact that $\sum_k c_k$ diverges for
$a<1$ does not imply anything about the convergence of~\eqref{W0-res}.  The issue
about the convergence of~\eqref{W0-res} thus remains open for $0<a\leq 1$.  
When $t=0$, however, something more can be said:
\begin{equation*}
	W_0(0,0) = \frac{1}{\pi} \sum_{k\geq 0} \frac{(1/2-a)_k}{(1/2)_k} = 
	{}_2F_1\left(\atop{1/2-a, 1}{1/2};1\right).
\end{equation*}
A Gauss hypergeometric series 
\begin{equation*}
	{}_2F_1\left(\atop{a,b}{c};1\right)
\end{equation*}
converges if and only if $\text{Re}(c-a-b) > 0$~\cite{Slater},
which in this case amounts to $a>1$.

The convergence behaviour of $W_1(p,q)$ is now easy to determine from~\eqref{W1}.
It is immediately clear that $W_1(p,q)$ is absolutely convergent for $a>0$.

For the convergence of~\eqref{A29} we follow the same strategy.  First, it 
is clear that the convergence of 
\begin{equation}
\sum_k b_k = \sum_{k\geq 2j} \frac{k!\,(1/2-a-2n)_k}{(1/2)_k(k-2j)!} L_k(p^2+q^2)
	\label{W2n-conv}
\end{equation}
is a sufficient condition for the convergence of~\eqref{A29} since the 
summation over $j$ is finite.  Then, we again use~\eqref{Lag-ineq}, and
in this case we get
\begin{equation*}
	|b_k| \leq \left|\frac{k!\,(1/2-a-2n)_k}{(1/2)_k(k-2j)!}\right| e^{t/2} = c_k.
\end{equation*}
Raabe's test yields in this case that the series $\sum_k c_k$ is convergent 
if $a>1+2j - 2n$.  Since $j \leq n$ (see the summation bounds in~\eqref{A29}), 
a sufficient (and necessary) condition for this 
to be true is $a>1$.  For $t=0$, use of the convergence of the Gauss hypergeometric
series for $z=1$ again yields convergence if and only if $a>1$.  Finally, for the 
functions $W_{2n+1}(p,q)$, we will thus have (absolute) convergence everywhere 
for $a>0$.

\end{document}